\documentclass[11pt,aps,prd,preprint,nofootinbib,superscriptaddress]{revtex4-1}
\usepackage{amsmath}
\usepackage[dvips]{graphicx}
\usepackage[english]{babel}
\newcommand{\be}{\begin{equation}}
\newcommand{\ee}{\end{equation}}
\newcommand{\ben}{\begin{eqnarray}}
\newcommand{\een}{\end{eqnarray}}

\begin{document}
\title{Thermodynamics of nonsingular bouncing universes}
\author{Pedro C. Ferreira\footnote{E-mail: pedro.ferreira@ect.ufrn.br}}
\affiliation{Escola de Ci\^{e}ncias e Tecnologia, Universidade
Federal do Rio Grande do Norte, Natal, Rio Grande do Norte,
59072-970, Brazil.}
\author{Diego Pav\'{o}n\footnote{E-mail: diego.pavon@uab.es}}
\affiliation{Departamento de F\'{\i}sica, Universidad Aut\'{o}noma
de Barcelona, 08193 Bellaterra (Barcelona), Spain.}

\begin{abstract}
\noindent Homogeneous and isotropic, nonsingular, bouncing world
models are designed to evade the initial singularity at the
beginning of the cosmic expansion. Here, we study the
thermodynamics of the subset of these models governed by general
relativity. Considering the entropy of matter and radiation and
considering the entropy of the apparent horizon to be proportional
to its area, we argue that these models do not respect the
generalized second law of thermodynamics, also away from the
bounce.
\end{abstract}
\maketitle


\section{Introduction}
\noindent The well-know, and partially successful, cosmological
standard model, based on  Einstein gravity, assumes a homogeneous
and isotropic universe whose energy density is supplied by matter
and fields that satisfy the strong energy condition
\cite{peebles}. The popular Einstein and de Sitter cosmology
\cite{edes} is a case in point. In this class of models an
inevitable singularity occurs at the beginning of the expansion.
As $t \rightarrow 0$ the energy density grows beyond whatever
bound, world lines cannot be continued any longer and, worse of
all, the spacetime description breaks down. This holds also true
if at very early times, before the conventional radiation
dominated phase, the expansion of the universe is driven by some
field whose energy density, essentially constant, overwhelms every
other form of energy \cite{inflation}.
\\  \

\begin{figure}[!htb]
  \begin{center}
    \begin{tabular}{c}
      \resizebox{90mm}{!}{\includegraphics{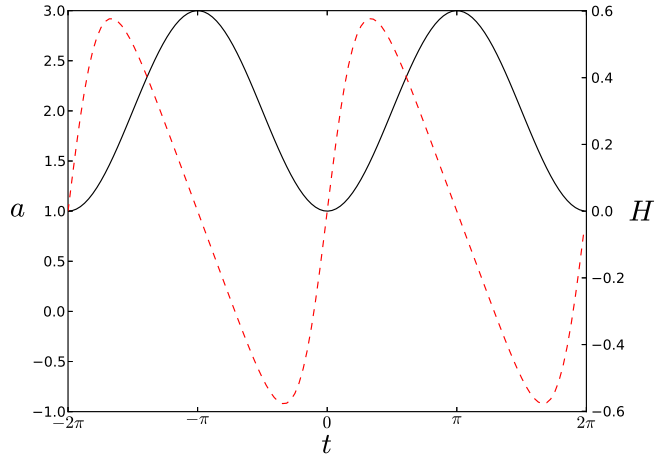}}\\
    \end{tabular}
    \caption{Qualitative evolution of the scale factor, solid (black) line, and the
    Hubble function, dashed (red) line, in arbitrary units, of a
    nonsingular bouncing FRW universe. Here, $a = 2 \, + \, \sin(t)$.}
    \label{fig:bounce}
 \end{center}
\end{figure}

\noindent To evade this uncomfortable, philosophical, implication
three main possibilities were advanced: (i) The initial
singularity arises chiefly from ignoring quantum effects which are
to prevail at some point before the singularity can occur
\cite{cai}. (ii) The universe began expanding from a finite
initial size safely larger than  Planck's length (thus avoiding as
well the quantum dominated regime). In this scenario, known as the
``emergent" universe, an Einstein-type static phase replaces the
initial singularity  \textemdash see e.g.
\cite{emergent1,emergent2}. (iii) Nonsingular bouncing universes.
According to this proposal the universe experiences either just
one contracting phase followed by an expanding one, or a sequence
of expanding and contracting phases connected, in either case, to
each other by a bounce such that the scale factor, $a(t)$, of the
Friedmann-Robertson-Walker (FRW) metric neither vanishes nor
diverges \cite{aleksei1978}-\cite{boisseau2015b}  \textemdash see
\cite{mario-review} and \cite{ppeter-review} for reviews. Since
the singularity theorems of Penrose and Hawking
\cite{royalsociety} forbid bounces to originate from normal matter
the physics behind them is not straightforward and requires the
presence of some source of energy that violates the null energy
condition. It is therefore understandable that the bounce
mechanism remains a subject of debate  as yet
\cite{rbrandenberger}. Nevertheless, nonsingular bouncing may
arise in de Sitter-like models with closed spatial sections
\cite{jerome-patrick} as well as in supergravity-based models
\cite{kohen2014}. Figure \ref{fig:bounce} shows the evolution of
the scale factor and Hubble factor, $H = \dot{a}/a$ \textemdash in
arbitrary units\textemdash  of a nonsingular, bouncing toy model.
In spite of its simplicity, it encapsulates the salient kinematics
features of this class of models.

\noindent Given the strong connection between gravity and
thermodynamics \cite{jakob1973}-\cite{pad2005}, for a cosmological
model that fits reasonably well the observational data it will be
a bonus to comply with the laws of thermodynamics, especially the
second law. The latter formalizes our daily experience that
macroscopic systems, left to themselves, spontaneously tend to
thermodynamic equilibrium or remain in it, if undisturbed.
According to it, the entropy $S$ of isolated systems never
decreases, $\dot{S} \geq 0$, and must be concave at least in the
last stage of approaching equilibrium \cite{callen1960}. In the
mid-1970s this was proved for gravity dominated systems, namely,
black holes plus their environment \cite{jakob1,jakob2}, and later
extended to cosmic scenarios consisting in a de Sitter event
horizon plus the fluid enclosed by it \cite{paul,diego}.
\\  \

\noindent In the case of  homogeneous and isotropic universes, the
entropy measured by a comoving observer splits into two parts, $S
= S_{\cal A}\, + \, S_{f}$. The first one is the entropy of the
apparent horizon \cite{bak-rey}
\begin{equation}
S_{\cal A} = k_{B} {\cal A}/(4\ell^{2}_{pl}) \;,
\label{eq:Sa}
\end{equation}
where ${\cal A} = 4 \pi \tilde{r}^{2}_{\cal A}$ denotes its area
and
\begin{equation}
\tilde{r}_{\cal A} = 1/\sqrt{H^{2} \, + \, k \, a^{-2}}
\label{eq:radius}
\end{equation}
denotes the horizon radius \cite{bak-rey}. This is in keeping with
the fact that the entropy of systems dominated by gravity does not
vary as its volume but as the area of the surface that bounds this
volume\footnote{This is, e.g., the case of the entropy of black
holes.}. The second part, $S_{f}$, corresponds to the entropy of
fluids (e.g. matter and radiation) enclosed by the horizon. In
many instances of interest the latter is negligible as compared
with the former; this is only natural given the smallness of the
$\ell^{2}_{pl}$ factor. (We do not consider the entropy of scalar
fields as we assume these fields  are in a pure quantum state and
consequently their entropy vanishes altogether). In the particular
case of the present universe the entropy of the horizon is larger
than that provided by supermassive black holes, stellar black
holes, relic neutrinos and CMB photons by $18$, $25$, $33$ and
$33$ orders of magnitude, respectively \cite{apj-egan}.
\\  \

\noindent At this point, we feel it expedient to summarize the
derivation of the radius of the apparent horizon, Eq.
(\ref{eq:radius}). We begin by rewriting the FRW metric,
\begin{equation}
{\rm d}s^{2} = -{\rm d}t^{2} \, + \, a^{2}(t)\, \frac{{\rm
d}r^{2}}{1-k r^{2}}\, + a^{2} (t) \, r^{2}\, {\rm d}\Omega^{2}
 \label{eq:frwmetric1}
\end{equation}
(where $k = +1, 0, -1$, denotes the spatial curvature index and
$\Omega$ is the unit two-sphere), as
\begin{equation}
{\rm d}s^{2} = h_{ab} {\rm d} x^{a}\, {\rm d} x^{b}\, + \, \tilde{
r}^{2}(x)\, {\rm d} \Omega^{2} \, . \label{eq:frwmetric2}
\end{equation}
Here, $x^{0} = t$, $x^{1} = r$, and $h_{ab} = {\rm diag} \left[
-1, \frac{a^{2}}{1-k r^{2}}\right]$. The radius of the dynamical
apparent horizon is set by the condition $\, h^{ab}
\partial_{a} \tilde{r} \, \partial_{b} \tilde{r} =0$. A
straightforward calculation produces Eq. (\ref{eq:radius}). To
better understand the meaning of the horizon note that the
expansion of the ingoing and outgoing null geodesic congruences
are given by
\begin{equation}
\theta_{IN} = H - \frac{1}{\tilde{r}}\, \sqrt{1 \, - \, \frac{k
\tilde{r}^{2}}{a^{2}}} \qquad {\rm and} \qquad \theta_{OUT} = H +
\frac{1}{\tilde{r}}\, \sqrt{1 \, - \, \frac{k
\tilde{r}^{2}}{a^{2}}}\, , \label{eq:inout}
\end{equation}
respectively. A spherically symmetric spacetime region will be
called a ``trapped" region if the expansion of ingoing and
outgoing null geodesics,  normal to the spatial two-sphere of
radius $\tilde{r}$ centered at the origin, is negative. By
contrast, the region will be called ``anti-trapped" if the
expansion of both kind of geodesics is positive. In normal regions
outgoing null rays have positive expansion and  ingoing null rays,
negative expansion. Thus, the anti-trapped region is given by the
condition $\tilde{r} > (H^{2} + k a^{-2})^{-1/2}$. Clearly, the
surface of the apparent horizon is nothing but the boundary
hypersurface of the  spacetime anti-trapped region. Therefore, it
is only natural to associate an entropy to the apparent horizon
since one can regard the latter as a measure of one's lack of
knowledge about the rest of the universe. Further details
concerning the entropy of the apparent horizon can be found in
\cite{bak-rey}. Obviously, in the case of an exact de Sitter
expansion, the apparent and event horizons coincide.

\noindent Models falling into the emergent class fulfill the
generalized  second law (GSL) \cite{plb2012}, i.e., $\dot{S}_{\cal
A}+ \dot{S}_{f} \geq 0$, as well as the concordance $\Lambda$CDM
model \cite{grg-ninfa1,ijgmmp-diego}. Likewise, the models of
references \cite{prd-ademir1} and \cite{mnras-ademir} satisfy the
said law \cite{prd-mimoso1}. Both scenarios evade the big bang
singularity and evolve from an initial de Sitter expansion to a
final, also de Sitter,  expansion in the far future. In the first
one, \cite{prd-ademir1}, a gravitationally induced production of
particles drives the current cosmic acceleration. In the second
one, \cite{mnras-ademir}, the latter arises from a dynamical
cosmological ``constant" that depends on the Hubble factor in a
manner dictated by the renormalization group \cite{jsola1}. By
contrast, ever expanding matter-dominated universes (as the
Einstein and de Sitter \cite{edes}) and phantom-dominated
cosmologies obeying Einstein gravity fail to satisfy this law
\cite{grg-ninfa1,ijgmmp-diego}.
\\  \

\noindent The target of this paper is to explore whether the GSL
is satisfied by the nonsingular bouncing class of world models
that obey general relativity, in their evolution outside the
bounce itself. As mentioned above, during the bounce the null
energy condition, $\rho \, + \, p \geq 0$, is violated;
consequently the second law is expected to be  violated as well.
It is therefore worthwhile to ask whether the transgression of the
second law is confined just to the bounce or it also occurs in
other phases of the universe evolution. This is the question we
aim to answer.
\\  \

\noindent Next section focuses on a homogeneous and  isotropic
cosmological model, sourcered  by a massive, complex,  scalar
field, that display a nonsingunlar bounce \cite{aleksei1978}.
Section III studies from an overall viewpoint the class of models
that incorporate matter and radiation as sources of the
gravitational field though, regrettably, these models do not have
analytical solutions. Section IV presents our conclusions and
final comments.

\section{Nonsingular Starobinsky's model}
\noindent Let us consider a model proposed by Starobinsky whose
only energy component is a homogeneous and massive, complex,
scalar field \cite{aleksei1978}. This is the simplest model
featuring a nonsingular bounce that has analytical solutions for
the scale factor with a nonzero measure in the space of initial
conditions  \textemdash in this last respect; see also
\cite{gibbons-turok2008}.

\noindent The model obeys  Friedmann's and energy conservation
 equations
\begin{equation}
H^{2}\, + \, \frac{k}{a^{2}} = \frac{8 \pi
G}{3}(|\dot{\varphi}|^{2}\, + \, m^{2} |\varphi|^{2})\, ,
\label{eq:alekfriedmann}
\end{equation}
\begin{equation}
\ddot{\varphi} \, + \, 3 H \dot{\varphi} \, + \, m^{2} \varphi = 0
\, ,
\label{eq:alekcontinuity}
\end{equation}
where  $m$ is the quantum mass of the field and $\, k = +1$  in
order for Eq. (\ref{eq:alekfriedmann}) to allow a bounce.
\\  \

\noindent At the epoch of maximum expansion ($-a_{\rm max} \ll t
\ll -m^{-1}$) the universe contracts according to $a(t) = \lambda
|t|^{3/2}$, where $\lambda$ is a positive-definite constant. In
this era the variation with time of the area of the apparent
horizon is
\begin{equation}
\dot{\cal A} = - 4 \pi \frac{\frac{8}{9}|t|^{-3}\, + \,
\frac{4}{3} \lambda^{-2} |t|^{-7/3}}{\left(
 \frac{4}{9}|t|^{-2}\, + \, \lambda^{-2} |t|^{-4/3}\right)^{2}} <
 0\, .
\label{eq:dotastarobisnky1}
\end{equation}
\\  \

\begin{figure}[!htb]
  \begin{center}
    \begin{tabular}{c}
      \resizebox{90mm}{!}{\includegraphics{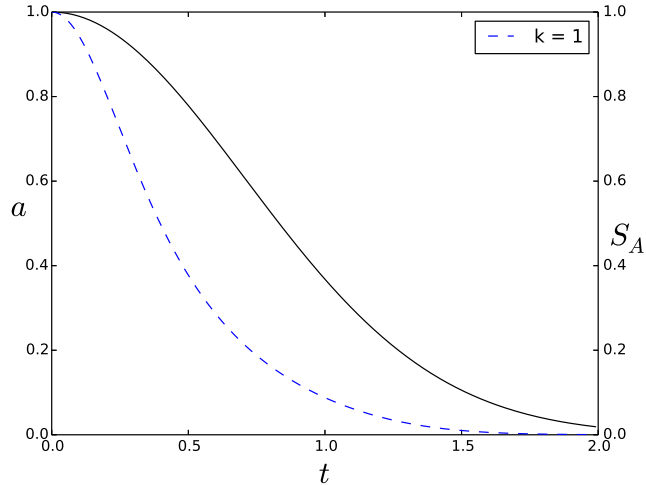}}\\
    \end{tabular}
    \caption{Evolution of the scale factor  \textemdash solid (black) line \textemdash  given by
    $ a(t) = a_{1} \, \exp(-C \, t^{2})$, and the entropy of the horizon
     \textemdash dashed (blue) line \textemdash for $t > 0$.}
    \label{fig:aleksei1978}
 \end{center}
\end{figure}

\noindent For $ t \gg m^{-1}$ but  $t <  m^{-1} (2 \ln m a_{\rm
max})^{1/2}$ the scale factor obeys $ a(t) = a_{1} \, \exp(-C \,
t^{2})$ with $C = m^{2}/6$ \cite{aleksei1978}. Consequently,
\begin{equation}
\dot{\cal A} = - 4 \pi \, t \, \frac{8 C^{2}\, + \, \frac{4C
\exp{(2C t^{2})}}{a_{1}^{2}}}{\left(4 C^{2} t^{2} \, + \,
\frac{\exp{(2C t^{2})}}{a_{1}^{2}}  \right)^{2}} < 0 \, .
\label{eq:dotastarobisnky2}
\end{equation}
Figure \ref{fig:aleksei1978} shows the evolution of the scale
factor (though, in reality,  the solution is valid only for  a
portion of the graph) as well as the evolution of the entropy of
the apparent horizon.

\noindent As we see,  the entropy of the horizon diminishes, at
least, in some interval of the contracting stage. Since neither
matter nor radiation sources the gravitational field the total
entropy decreases in such periods.

\section{Models with matter and radiation}
\begin{figure}[!htb]
  \begin{center}
    \begin{tabular}{c}
      \resizebox{90mm}{!}{\includegraphics{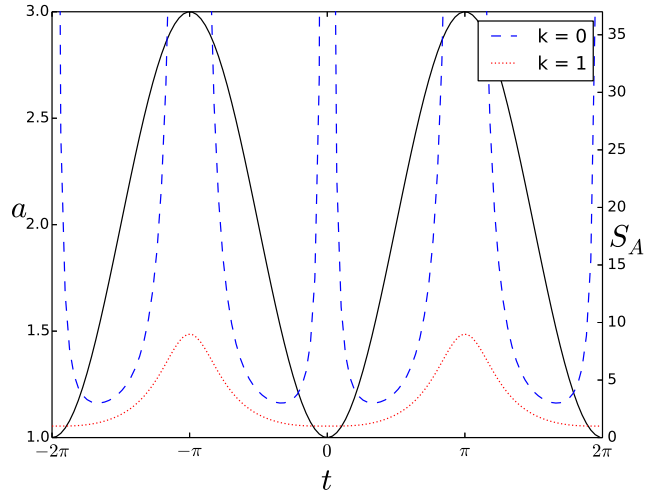}}\\
    \end{tabular}
    \caption{Evolution of the scale factor \textemdash solid (black)
    line \textemdash  and the horizon entropy, in arbitrary units,
    for $k = 0$  \textemdash dashed (blue) lines \textemdash, and
    $ k= +1$ \textemdash dotted (red) line. Here,
    $a = 2 + \,  \sin(t)$.}
    \label{fig:oscillations1}
 \end{center}
\end{figure}

\noindent We next consider nonsingular bouncing models that
include matter and radiation. Since we are not aware of analytical
solutions for the scale factor our study will necessarily be from
a general viewpoint. In the following we adopt the toy model of
Fig. \ref{fig:oscillations1}.
\\   \

\noindent Inspection of the said figure reveals that irrespective
of the sign of the curvature index, $k$, of the FRW metric
(\ref{eq:frwmetric1}), the entropy of the apparent horizon
$S_{\cal A}$, oscillates so that it presents alternated phases of
increase and decrease. To fulfill the GSL, the entropy of matter
and fields inside the horizon should compensate for the stages of
decrease. Even though this can hardly be the case \footnote{Note
that, as mentioned above, in most instances, the entropy of the
horizon overwhelms any other form of entropy \cite{apj-egan}.},
this may not be necessarily true near the turning points (where
$\dot{a}$ vanishes). In these points, it may depend on the details
of the specific bouncing model. So, except possibly for the
turning points, the entropy of matter and fields inside the
horizon, most likely,  will not compensate.
\\  \

\noindent In Fig. \ref{fig:oscillations1} it is seen that when the
scale factor goes from a maximum to the next minimum, the area,
${\cal A} = 4 \pi /(H^{2}\, + \, k a^{-2})$ (or what amounts to
the same, the entropy), of the $k = + 1$ apparent horizon
decreases. This is also true for the $k = 0$ apparent horizon when
the scale factor goes from a maximum to the consecutive inflection
point. Next we shall check if matter and radiation may compensate
for the decrease in entropy.
\\  \

\noindent The entropy of pressureless matter, as well as of
massless radiation, inside the apparent horizon varies as the
number of particles in there, i.e., $S_{m} = k_{B} N_{m}$,
$S_{\gamma} = k_{B} N_{\gamma}$ \cite{Frautschi}. Since $N_{i} =
(4 \pi/3) \, n_{i} \, \tilde{r}_{\cal A}^{3}$, where the number
densities obey $n_{m} = n_{m0} \, a^{-3}$ and $n_{\gamma} =
n_{\gamma 0} \, a^{-4}$ for matter and radiation,
respectively\footnote{For simplicity we have set $a_{0} =1$.}, one
obtains
\begin{equation}
\dot{S}_{m} = - 4 \pi k_{B} \, n_{m0} \,
 \left(\frac{\tilde{r}_{\cal A}}{a} \right)^{5}\, \dot{a} \, \ddot{a} \, ,
 \label{eq:dotsmatter}
\end{equation}
and
\begin{equation}
\dot{S}_{\gamma} = - 4 \pi k_{B} \, n_{\gamma 0} \,
 \left(\frac{\tilde{r}_{\cal A}}{a} \right)^{3}\, H\,
 \left[\left(\frac{\tilde{r}_{\cal A}}{a}\right)^{2} \, \ddot{a} + \,
 \frac{1}{3a}\right].
 \label{eq:dotsradiation}
\end{equation}
Inspection of the right-hand-side of the last two equations and
Fig. \ref{fig:oscillations1}, reveals that the entropy of the
fluids (matter and radiation)  enclosed by the apparent horizon
also diminishes in the scale factor interval between each maximum
and the subsequent inflection point (note that there both
$\dot{a}$ and $\ddot{a}$ are negative quantities). Thereby, in the
said interval the total entropy (that of the horizon plus that of
relativistic and non-relativistic particles inside the horizon)
decreases, $\dot{S}_{\cal A}\, + \dot{S}_{m} \, +\,
\dot{S}_{\gamma} < 0$.
\\  \

\noindent In addition, as is well known, when matter evolves from
an uniform distribution to a clumped one to form bound objects
(stars, galaxies, and so on) its entropy also diminishes since
spatial matter distribution goes from a disordered state to a less
disordered one. In expanding phases of an oscillating universe
this decrease might be counterbalanced by the entropy gain
associated to the volume increase of the universe, but not in
contracting phases.
\\  \

\noindent Likewise, if one includes the ``entropy" associated to
the perturbed FRW universe, recently introduced by Clifton et al.
\cite{cqg-clifton}, based on the Bel-Robinson tensor, the
situation does not change since the said entropy is  proportional
to $a^{3}$ (see Eq. (54) in \cite{cqg-clifton}). Then in
contracting phases this entropy diminishes as well.
\\  \

\noindent For a bounce (or a sequence of bounces) to occur in the
class of models considered in this section, an energy component
that violates the null energy condition (NEC) must also be
present. Thus, one may wonder if this component may restore the
GSL in this context. Typically, fields that break the NEC are
scalar fields of phantom type. As mentioned in the Introduction,
it is generally assumed that scalar fields have vanishing entropy.
Upon this usually accepted assumption, the presence of this field
will not modify our results in any way. However, one may still
argue that the said field might, in reality, be an effective one
that corresponds to a mixture of fields,  every single one being
in a pure quantum state. In such a case, this effective phantom
field should be entitled to an entropy. Now, the entropy of any
system is an increasing function of its energy and volume. In the
contracting phase, at variance with matter and fields that fulfill
the NEC, the energy density of the phantom field decreases. Since
the volume of the apparent horizon diminishes as well, the entropy
of this effective, phantom field, decreases too. Once again, our
result stands.
\\  \

\noindent Altogether, our analysis suggests that the GSL is not
satisfied by homogeneous and isotropic, nonsingular, bouncing
models obeying Einstein relativity.

\section{Concluding remarks}
\noindent Assuming that the entropy of the universe is
proportional to the area enclosing the dynamical apparent horizon,
and taking into account the contributions by the matter and
radiation, we investigated if nonsingular bouncing models satisfy
the generalized second law of thermodynamics. We found that
homogeneous and isotropic nonsingular bouncing models, governed by
general relativity, do not comply with the GSL also away from the
bounce. This is not only an interesting result in itself, it also
begs the question as to whether more elaborated bouncing models
respect this law.
\\  \

\noindent One may wonder whether our results would change if we
have used the event horizon rather than the apparent horizon. The
answer is: no. The entropy of the former horizon is also given by
his area, ${\cal A}_{eh} \propto l^{2}_{eh}$ where the radius of
the horizon reads $l_{eh}(t) = a(t)\, \int_{t}^{\infty}{{\rm
d}\tilde{t}/a(\tilde{t})}$. Consequently, in contracting phases
this entropy decreases as well.
\\  \

\noindent Similarly, the ``entropy of the gravitational field"
introduced four decades ago by Penrose \cite{penrose1,penrose2},
based on Weyl's tensor does not change our results because, in the
case of non-perturbed FRW universes, the Weyl tensor vanishes
identically and so does this entropy. At any rate, it faces well
known difficulties; see e.g. \cite{bonnor,goode}.
\\  \

\noindent One might still argue that the failure to fulfill the
GSL arises from the fact that we are observing just one universe
and that we did not take into account cosmic variance. However,
from the principle of indifference of  Laplace \cite{Laplace}, it
is hard to see how cosmic variance will alter this result for the
subset of the multiverse\footnote{By multiverse we mean the set of
all model universes.} obeying general relativity. In any case,
this depends on how one would choose a measure in the space of
solutions, a subject that lies beyond the scope of this work.
\\  \

\noindent Finally, bouncing models based on gravity theories that
go beyond general relativity, or that introduce extra dimensions
(see \cite{mario-review,ppeter-review} and references therein, as
well as the recent proposals by Steinhardt et al.\cite{steinhardt,
bars} and Ijjas and Steinhardt \cite{ijjas}), should also be
considered from the thermodynamic standpoint. However, for the
time being, it is rather unclear how to define a meaningful
entropy in such models.
\\  \

\noindent We conclude by saying that, while the aforesaid
connection between gravity and thermodynamics,
\cite{jakob1973}-\cite{pad2005}, does not fully ensure us that the
universe behaves as a thermodynamic system (i.e., one that obeys
the thermodynamic laws), given two models that  fit equally well
the observational data,  one respecting the GSL while the other
does not, the former will likely be preferred.

\acknowledgments{\noindent We thank the anonymous referee for
useful suggestions and advice. PCF acknowledges financial support
from CNPq (Brazil) and is grateful to the Department of Physics of
the ``Universidad Aut\'{o}noma de Barcelona" (where part of this
work was done) for warm hospitality. This research was partly
supported by the ``Ministerio de Econom\'{\i}a y Competitividad,
Direcci\'{o}n General de Investigaci\'{o}n Cient\'{\i}fica y
T\'{e}cnica", Grant No. FiS2012-32099.}


\begin{thebibliography}{99}
\bibitem{peebles} P.J.E. Peebles, {\em Principles of Physical
Cosmology} (Princeton University Press, Princeton,  1993)
\bibitem{edes} A. Einstein and W. de Sitter, Proc. Natl. Acad.
Sci. \underline{18}, 214 (1932)
\bibitem{inflation} A. Linde, {\em Particle Physics and Inflationary
Cosmology} (Harwood Academic, Chur,  1990)
\bibitem{cai} Y.-F. Cai and E. Wilson-Ewing, J. Cosmol. Astropart.
Phys. 03 (2015) 006
\bibitem{emergent1} G.F.R. Ellis and R. Maartens, Class. Quantum
Grav. \underline{21}, 223 (2004)
\bibitem{emergent2} G.F.R. Ellis, J. Murugan, C.G. Tsagas,
Class. Quantum Grav. \underline{21}, 233 (2004)
\bibitem{aleksei1978} A.A. Starobinsky, Sov. Astron. Lett.
\underline{4}, 82 (1978)
\bibitem{kamenshchik1997} A.Y. Kamenenshchik, I.M. Khalatnikov and
A.V. Toporensky, Int. J. Mod. Phys. D \underline{7}, 673 (1997).
\bibitem{rbrandenberger} R. Brandenberger, AIP Conf. Proc.
\underline{1268}, 3 (2010)
\bibitem{bamba} K. Bamba {\it et al.}, J. Cosmol. Astropart. Phys.
\underline{04}, 001 (2015)
\bibitem{boisseau2015a} B. Boisseau, H. Giacomini, D. Polarski, and  A.A.
Starobinsky, J. Cosmol. Astropart. Phys. \underline{07}, 002
(2015)
\bibitem{boisseau2015b} B. Boisseau, H. Giacomini, and D. Polarski, J. Cosmol.
Astropart. Phys. \underline{10}, 033 (2015)
\bibitem{mario-review} M. Novello and S.E. Perez Bergliaffa,
Phys. Rep. \underline{463}, 127 (2008)
\bibitem{ppeter-review} D. Battefeld and P. Peter,
Phys. Rep. \underline{571}, 1 (2015)
\bibitem{royalsociety} R. Penrose and S.W. Hawking, Proc. Roy.
Soc. London \underline{A314}, 529 (1970)
\bibitem{jerome-patrick} J. Martin and P. Peter, Phys. Rev. D
\underline{68}, 103517  (2003)
\bibitem{kohen2014} M. Kohen, J.L. Lehners, and B.A. Ovrut, Phys.
Rev. D \underline{90}, 025005 (2014)
\bibitem{jakob1973} J.D. Bekenstein, Phys. Rev. D \underline{7},
 2333 (1973)
\bibitem{gary1977} G. Gibbons and S.W. Hawking, Phys. Rev. D
\underline{15}, 2738  (1977)
\bibitem{Jacosbson1995} T. Jacobson, Phys. Rev. Lett.
\underline{75}, 1260 (1995)
\bibitem{pad2005} T. Padmanabhan, Phys. Rep. \underline{406},
 49 (2005)
\bibitem{callen1960} H. Callen, {\em Thermodynamics} (Wiley, NY,
1960)
\bibitem{jakob1} J.D. Bekenstein, Phys. Rev. D \underline{9},  3292 (1974)
\bibitem{jakob2} J.D. Bekenstein, Phys. Rev. D \underline{12}, 3077 (1975)
\bibitem{paul} P.C.W. Davies, Class. Quantum Grav. \underline{4}, L225
(1987).
\bibitem{diego} D. Pav\'{o}n, Class. Quantum Grav. \underline{7}, 487
(1990).
\bibitem{bak-rey} D. Bak and S.J. Rey, Class. Quantum Grav.
\underline{17}, L83  (2000)
\bibitem{apj-egan} C.A. Egan and C.H. Lineweaver,
Astrophys. J. \underline{710}, 1825  (2010)
\bibitem{plb2012} S. del Campo, R.  Herrera, and
D. Pav\'{o}n, Phys. Lett. B \underline{707}, 8 (2012)
\bibitem{grg-ninfa1} N. Radicella and D. Pav\'{o}n, Gen.
Relativ. Grav. \underline{44}, 685  (2012)
\bibitem{ijgmmp-diego} D. Pav\'{o}n, Int. J. Geom. Methods Mod. Phys.
 \underline{11}, 1460007 (2014)
\bibitem{prd-ademir1} J.A.S. Lima, S. Basilakos, and F.E.M. Costa,
Phys. Rev. D \underline{86}, 1035034  (2012)
\bibitem{mnras-ademir} J.A.S. Lima, S. Basilakos, and J. Sol\'{a},
Mon. Not. R. Astron. Soc. \underline{431}, 923 (2013)
\bibitem{prd-mimoso1} J.P. Mimoso and D. Pav\'{o}n,
Phys. Rev. D \underline{87}, 047302 (2013)
\bibitem{jsola1} J. Sol\'{a}, J. Phys. Conf. Ser.
\underline{283}, 012033 (2011)
\bibitem{gibbons-turok2008} G.W. Gibbons and N. Turok, Phys. Rev.
D \underline{77}, 063516 (2008)
\bibitem{Frautschi} S. Frautschi, Science \underline{217},
13 (1982)
\bibitem{cqg-clifton} T. Clifton, G.F.R. Ellis and  R. Tavakol,
Class. Quantum  Grav. \underline{30}, 125009 (2013)
\bibitem{penrose1} R. Penrose, in {\em Proceedings of the First Marcel Grossmann
Meeting on General Relativity}, ed. R. Ruffini (ICTP, Triestre,
North Holland, Amsterdam, 1977)
\bibitem{penrose2} R. Penrose, in {\em General Relativity: An Einstein Centenary
Survey}, ed.  by S.W. Hawking, W. Israel (Cambridge University
Press, Cambridge, 1979)
\bibitem{bonnor} W.B. Bonnor, Phys. Lett. A \underline{122},
305 (1987)
\bibitem{goode} S.W. Goode, A.A. Coley, and J. Wainwright, Class. Quantum Grav
\underline{9}, 445 (1992)
\bibitem{Laplace} P.S. Laplace, in {\em A Philosophical Essay on Probabilities}, ed. by
F.W. Truscott, F.L. Emory (Dover, New York, 1951)
\bibitem{steinhardt} P.J. Steinhardt and N. Turok, Science
\underline{296}, 1436 (2002)
\bibitem{bars} I. Bars, P. Steinhardt, and N. Turok, Phys. Rev. D
\underline{89}, 043515 (2014)
\bibitem{ijjas} A. Ijjas and P. Steinhardt, J. Cosmol. Astropart. Phys.
\underline{10}, 001 (2015)
\end{thebibliography}
\end{document}